\documentclass[preprint,aps,12pt,showpacs,nofootinbib,tightenlines]{revtex4}
\usepackage{amsmath}
\usepackage{amssymb}
\usepackage{epsfig}
\usepackage{graphicx}
\def\be{\begin{eqnarray}}
\def\en{\end{eqnarray}}
\def\non{\nonumber\\}

\def\ra{\rangle}

\def\sl{\!\!\!\slash}
\def\prd{{Phys. Rev. D}~}
\def\prl{{ Phys. Rev. Lett.}~}
\def\plb{{ Phys. Lett. B}~}

\def\epjc{{ Eur. Phys. J. C}~}
\newcommand{\acp}{{\cal A}_{CP}}

\newcommand{\etapp}{\eta^{(\prime)}}
\begin{document}
\title{Study of scalar meson $a_0(980)$ from $B \to a_0(980)\pi$ Decays }
\author{ Zhi-Qing
Zhang$^{a,b}$\footnote{Electronic address: zhangzhiqing@zzu.edu.cn} and Zhen-jun Xiao$^b$
\footnote{Electronic address: xiaozhenjun@njnu.edu.cn}} \affiliation{
{\it \small $a$ Department of Physics, Henan University of Technology,
Zhengzhou, Henan 450052, P.R.China;}\\
{\it \small $b$ Department of Physics and Institute of
Theoretical Physics, Nanjing Normal University, Nanjing, Jiangsu
210097, P.R.China}} 
\begin{abstract}
\date{\today}

In this paper, we calculate the branching ratios and the direct CP-violating
asymmetries for decays $\bar{B}^0\to a^0_0(980)\pi^0, a^+_0(980)\pi^-,
a^-_0(980)\pi^+$ and $B^-\to a^0_0(980)\pi^-, a^-_0(980)\pi^0$
by employing the perturbative QCD factorization approach.
Although the light scalar meson $a_0(980)$ is widely perceived as
primarily the four-quark bound states, the calculation shows that the 2-quark model
supposition for $a_0(980)$ can not be ruled out by the currently available experiment
upper limits.
In these considered decays, the branching ratio for the  $B^-\to a^0_0(980)\pi^-$
is the largest, $2.8\times 10^{-6}$, while its direct CP asymmetry is
the smallest, $\sim 14\%$.
Although the direct CP asymmetries for the decays $\bar{B}^0\to a^0_0(980)\pi^0
, B^-\to a^-_0(980)\pi^0$
are large, about $(70\thicksim80)\%$, it is still difficult to measure them,
since their branching ratios are small, around $(4\thicksim5)\times 10^{-7}$.

\end{abstract}

\pacs{13.25.Hw, 12.38.Bx, 14.40.Nd}
\vspace{1cm}

\maketitle


\section{Introduction}\label{intro}

The study about scalar meson is an interesting topic for both theory
and experiment. In order to uncover their mysterious structure,
intensive studies have been done for the B meson decays involving a
scalar meson as one of the two final state mesons. Such kind of
decays have been studied by employing various factorization
approaches, such as the generalized factorization approach
\cite{GMM}, the QCD factorization (QCDF) approach
\cite{CYf0K,CCYscalar}, the perturbative QCD (pQCD) approach
\cite{Chenf0K1,Chenf0K2,wwang,ylshen,zhangzq1}, and by using the QCD
sum rule \cite{maozhi1,maozhi2,maozhi3}.

On the experimental side, from the first scalar meson $f_0(980)$ observed
in the decay mode $B\to f_0(980)K$ by Belle \cite{Bellef0}, and
confirmed by BaBar \cite{BaBarf0} later, then many
channels involving a scalar in the final state have been measured by
Belle \cite{Belle1,Belle2} and BaBar \cite{BaBar3,BaBar4}.
For example, the decays $B\to a_0(980)\pi$ were searched by BarBar five years ago
\cite{BaBar5},
especially the decay $B^-\to a^-_0(980)\pi^0$, which has been considered as a
best candidate to distinguish the nature of the scalar $a_0(980)$ \cite{delipin}.
The authors of Ref.\cite{delipin}
argued that if the branching ratio of this channel can be measured accurately by
the experiment, one can separate the four- and two-quark assignments,
because the results of these two assignments
have a difference of one order of magnitude.
So in the past three years, BarBar have given this channel
twice measurements \cite{a0pi1,a0pi2} and get two almost identical upper limits.
For our considered decays, only the
experimental upper limits are available now for some of them \cite{pdg08}:
\be
Br(\bar{B}^0\to a^+_0(980)\pi^-) < 3.1\times 10^{-6},\non
Br(B^-\to a^-_0(980)\pi^0) < 1.4\times 10^{-6}, \non
Br(B^-\to a^0_0(980)\pi^-) < 5.8\times 10^{-6}.
\en

In this paper, we will study the branching ratios and
CP asymmetries of $\bar{B}^{0} \to a^0_0(980)\pi^{0}, a^\pm_0(980)\pi^\mp$ and $B^-\to
a^-_0(980)\pi^0, a^0_0(980)\pi^-$ within perturbative
QCD approach based on $k_T$ factorization.
In the following, we use $a_0$ to denote $a_0(980)$ in some places for convenience.
It is organized as follows. In Sect.\ref{proper}, the status of the
study on the physical properties of $a_0$, the relevant decay constants
and light-cone distribution amplitudes are  discussed.
In Sec.\ref{results}, we then analysis these decay channels using the pQCD approach.
The numerical results and the discussions are given
in the section \ref{numer}. The conclusions are presented in the final part.


\section{Physical Properties Of The Final Particles}\label{proper}

Many scalar mesons below 2GeV have been found in experiments. We cann't accommodate these scalar mensons into one nonet, but need at least two  nonets
below and above 1 GeV\cite{close}. Among them, the scalar mesons below 1 GeV, including
$f_0(600)(\sigma), f_0(980), K_0^*(800)(\kappa)$ and $a_0(980)$, are usually viewed to form an SU(3) nonet;
while scalar mseons above 1 GeV, including $f_0(1370), f_0(1500)/f_0(1700), K^*(1430)$ and $a_0(1450)$ form the
other SU(3) nonet.
There are several  different scenarios to describe these mesons in the
quark model \cite{nato,jaffe,jwei,baru}. For example $a_0(980)$ meson, which has been suggested as $\bar{q} q$ lowest lying state \cite{nato}(called scenario I)
or four-quark bound state \cite{jaffe}(called scenario II). In the scenario I, the former  SU(3) nonet mesons are treated as the
$\bar{q} q$ ground stats,
and the latter nonet ones are the first excited states; in the scenario II, the former nonet mesons
are viewed as four-quark bound states, while the latter nonet ones are $\bar{q} q$ ground states.
Some people also consider that it is not made of one simple component but might have a more complex nature such as having
a $K\bar{K}$ component\cite{jwei,baru},  even the superpositions of the two- and four- quark states.
In order to make quantitative
predictions, we identify $a_0(980)$ as the two-quark state in the calculation.

In  2-quark model, the decay constants for scalar meson $a_0$
are defined by:
\be
\langle a_0(p)|\bar q_2\gamma_\mu
q_1|0\ra&=&f_{a_0}p_\mu, \,\,\,\, \langle a_0(p)|\bar q_2
q_1|0\ra=m_{a_0}\bar {f}_{a_0}.
\en
For the neutral scalar meson $a_0$
cannot be produced via the vector current (restricted by the charge
conjugation invariance or the G parity conservation),
the vector decay constant $f_{a_0}=0$. As to the charged scalar mesons $a_0^-$, from the equation of motion:
\be
\mu_{a_0^-}f_{a_0^-}=\bar {f}_{a_0^-}, \quad\quad\quad\quad{\rm with} \quad \mu_{a_0^-}=\frac{m_{a_0^-}}{m_d(\mu)-m_u(\mu)},
\en
its vector decay constant is
proportional to the mass difference between the constituent $u$ and $d$ quarks. It is easy to see the vector decay
constant is very small, and will equal to zero in the SU(3) limit. So we only need consider the scalar decay constant
$\bar f_{a_0}$, which is scale dependent. Fixing the scale at 1 GeV, the value is $\bar {f}_{a_0}=(365\pm20)$MeV,
which is calculated in QCD sum rules\cite{CCYscalar}.

The light-cone distribution amplitudes (LCDAs) for the  scalar
meson $a_0$ can be written as:
\be \langle
a_0(p)|\bar q_1(z)_l q_2(0)_j|0\rangle
&=&\frac{1}{\sqrt{2N_c}}\int^1_0dx \; e^{ixp\cdot z}\non
&&
\times \{ p\sl\Phi_{a_0}(x)
+m_{a_0}\Phi^S_{a_0}(x)+m_{a_0}(n\sl_+n\sl_--1)\Phi^{T}_{a_0}(x)\}_{jl},\quad\quad\label{LCDA}
\en
where $n_+$ and $n_-$ are light-like vectors:
$n_+=(1,0,0_T),n_-=(0,1,0_T)$, and $n_+$ is parallel with the moving direction of the scalar meson $a_0$.
The normalization can be related to
the decay constants:
\be \int^1_0 dx\Phi_{a_0}(x)=\int^1_0
dx\Phi^{T}_{a_0}(x)=0,\,\,\,\,\,\,\,\int^1_0
dx\Phi^{S}_{a_0}(x)=\frac{\bar f_{a_0}}{2\sqrt{2N_c}}\;.
\en
The twist-2 LCDA can be expanded in the Gegenbauer polynomials:
\be
\Phi_{a_0}(x,\mu)&=&\frac{1}{2\sqrt{2N_c}}\bar
f_{a_0}(\mu)6x(1-x)\sum_{m=1}^\infty B_m(\mu)C^{3/2}_m(2x-1),
\en
the values for Gegenbauer moments $B_1, B_3$ have been calculated in \cite{CCYscalar} as:
\be
B_1=-0.93\pm0.10,\quad\quad B_3=0.14\pm0.08.
\en
These vaues are taken at $\mu=1$ GeV and the even Gegenbauer moments vanish.

As for the twist-3 distribution amplitudes $\Phi_{a_0}^s$ and $\Phi_{a_0}^T$,
they have not been studied in the literature, so we adopt the asymptotic form \cite{ylshen}:
\be
\Phi^S_{a_0}&=& \frac{1}{2\sqrt {2N_c}}\bar f_{a_0},\,\,\,\,\,\,\,\Phi_{a_0}^T=
\frac{1}{2\sqrt {2N_c}}\bar f_{a_0}(1-2x).
\en


\section{ The perturbative QCD  calculation} \label{results}

Under the two-quark model for the scalar meson $a_0$ supposition, we
would like to use pQCD approach to study B decays into $a_{0}$
and $\pi$.
The decay amplitude can be conceptually written as the convolution,
\be
{\cal A}(B \to \pi a_0)\sim \int\!\! d^4k_1
d^4k_2 d^4k_3\ \mathrm{Tr} \left [ C(t) \Phi_B(k_1) \Phi_{\pi}(k_2)
\Phi_{a_0}(k_3) H(k_1,k_2,k_3, t) \right ], \label{eq:con1}
\en
where $k_i$'s are momenta of light quarks included in each mesons, and
$\mathrm{Tr}$ denotes the trace over Dirac and color indices. $C(t)$
is the Wilson coefficient which results from the radiative
corrections at short distance. In the above convolution, $C(t)$
includes the harder dynamics at larger scale than $M_B$ scale and
describes the evolution of local $4$-Fermi operators from $m_W$ (the
$W$ boson mass) down to $t\sim\mathcal{O}(\sqrt{\bar{\Lambda} M_B})$
scale, where $\bar{\Lambda}\equiv M_B -m_b$. The function
$H(k_1,k_2,k_3,t)$ describes the four quark operator and the
spectator quark connected by
 a hard gluon whose $q^2$ is in the order
of $\bar{\Lambda} M_B$, and includes the
$\mathcal{O}(\sqrt{\bar{\Lambda} M_B})$ hard dynamics. Therefore,
this hard part $H$ can be perturbatively calculated. The function
$\Phi_{(\pi,a_0)}$ are the wave functions of $\pi$ and
 $a_0$, respectively.

Since the b quark is rather heavy we consider the $B$ meson at rest
for simplicity. It is convenient to use light-cone coordinate $(p^+,
p^-, {\bf p}_T)$ to describe the meson's momenta, \be p^\pm =
\frac{1}{\sqrt{2}} (p^0 \pm p^3), \quad {\rm and} \quad {\bf p}_T =
(p^1, p^2). \en Using these coordinates the $B$ meson and the two
final state meson momenta can be written as \be P_B =
\frac{M_B}{\sqrt{2}} (1,1,{\bf 0}_T), \quad P_{2} =
\frac{M_B}{\sqrt{2}}(1,0,{\bf 0}_T), \quad P_{3} =
\frac{M_B}{\sqrt{2}} (0,1,{\bf 0}_T), \en respectively. The meson
masses have been neglected. Putting the anti- quark momenta in $B$,
$P$ and $S$ mesons as $k_1$, $k_2$, and $k_3$, respectively, we can
choose
\be k_1 = (x_1 P_1^+,0,{\bf k}_{1T}), \quad k_2 = (x_2
P_2^+,0,{\bf k}_{2T}), \quad k_3 = (0, x_3 P_3^-,{\bf k}_{3T}). \en
For these considered decay channels, the integration over $k_1^-$,
$k_2^-$, and $k_3^+$ in eq.(\ref{eq:con1}) will lead to
\be
 {\cal
A}(B \to \pi a_0) &\sim &\int\!\! d x_1 d x_2 d x_3 b_1 d b_1 b_2 d
b_2 b_3 d b_3 \non && \cdot \mathrm{Tr} \left [ C(t) \Phi_B(x_1,b_1)
\Phi_{\pi}(x_2,b_2) \Phi_{a_0}(x_3, b_3) H(x_i, b_i, t) S_t(x_i)\,
e^{-S(t)} \right ], \quad\quad \label{eq:a2}
\en
where $b_i$ is the
conjugate space coordinate of $k_{iT}$, and $t$ is the largest
energy scale in function $H(x_i,b_i,t)$.
In order to smear the end-point singularity on $x_i$,
the jet function $S_t(x)$ \cite{li02}, which comes from the
resummation of the double logarithms $\ln^2x_i$, is used
\be
 S_t(x)=\frac{2^{1+2c}\Gamma(3/2+c)}{\sqrt{\pi}\Gamma(1+c)}[x(1-x)]^c,
\en
where the parameter $c=0.4$. The
last term $e^{-S(t)}$ in Eq.(\ref{eq:a2}) is the Sudakov form factor which suppresses
the soft dynamics effectively \cite{soft}.

 For the considered decays, the related weak effective
Hamiltonian $H_{eff}$ can be written as \cite{buras96}
\be
\label{eq:heff} {\cal H}_{eff} = \frac{G_{F}} {\sqrt{2}} \,
\sum_{q=u,c}V_{qb} V_{qd}^*\left[ \left (C_1(\mu) O_1^q(\mu) +
C_2(\mu) O_2^q(\mu) \right) \sum_{i=3}^{10} C_{i}(\mu) \,O_i(\mu)
\right] \;,
\en
with the Fermi constant $G_{F}=1.166 39\times
10^{-5} GeV^{-2}$, and the CKM matrix elements V. We specify below
the operators in ${\cal H}_{eff}$ for $b \to d$ transition: \be
\begin{array}{llllll}
O_1^{u} & = &  \bar d_\alpha\gamma^\mu L u_\beta\cdot \bar
u_\beta\gamma_\mu L b_\alpha\ , &O_2^{u} & = &\bar
d_\alpha\gamma^\mu L u_\alpha\cdot \bar
u_\beta\gamma_\mu L b_\beta\ , \\
O_3 & = & \bar d_\alpha\gamma^\mu L b_\alpha\cdot \sum_{q'}\bar
 q_\beta'\gamma_\mu L q_\beta'\ ,   &
O_4 & = & \bar d_\alpha\gamma^\mu L b_\beta\cdot \sum_{q'}\bar
q_\beta'\gamma_\mu L q_\alpha'\ , \\
O_5 & = & \bar d_\alpha\gamma^\mu L b_\alpha\cdot \sum_{q'}\bar
q_\beta'\gamma_\mu R q_\beta'\ ,   & O_6 & = & \bar
d_\alpha\gamma^\mu L b_\beta\cdot \sum_{q'}\bar
q_\beta'\gamma_\mu R q_\alpha'\ , \\
O_7 & = & \frac{3}{2}\bar d_\alpha\gamma^\mu L b_\alpha\cdot
\sum_{q'}e_{q'}\bar q_\beta'\gamma_\mu R q_\beta'\ ,   & O_8 & = &
\frac{3}{2}\bar d_\alpha\gamma^\mu L b_\beta\cdot
\sum_{q'}e_{q'}\bar q_\beta'\gamma_\mu R q_\alpha'\ , \\
O_9 & = & \frac{3}{2}\bar d_\alpha\gamma^\mu L b_\alpha\cdot
\sum_{q'}e_{q'}\bar q_\beta'\gamma_\mu L q_\beta'\ ,   & O_{10} & =
& \frac{3}{2}\bar d_\alpha\gamma^\mu L b_\beta\cdot
\sum_{q'}e_{q'}\bar q_\beta'\gamma_\mu L q_\alpha'\ ,
\label{eq:operators} \end{array}
\en
where $\alpha$ and $\beta$ are
the $SU(3)$ color indices; $L$ and $R$ are the left- and
right-handed projection operators with $L=(1 - \gamma_5)$, $R= (1 +
\gamma_5)$. The sum over $q'$ runs over the quark fields that are
active at the scale $\mu=O(m_b)$, i.e., $(q'\epsilon\{u,d,s,c,b\})$.

\begin{figure}[t,b]
\vspace{-3cm} \centerline{\epsfxsize=16 cm \epsffile{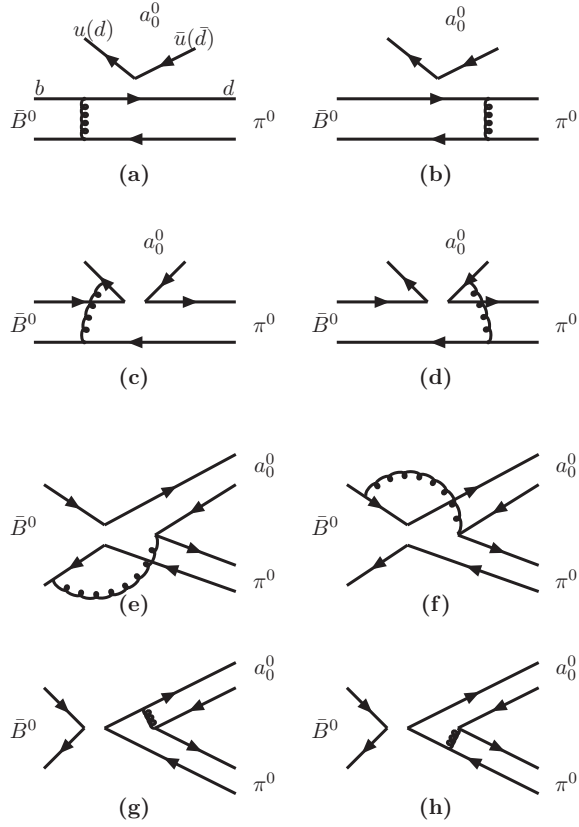}}
\vspace{-9cm} \caption{ Diagrams contributing to the decay $\bar{B}^0\to \pi^0
a^0_0$ .}
 \label{fig1}
\end{figure}

In the following, we take the $\bar{B}^0\to\pi^0 a_0^0$ decay
channel as an example to expound.
There are 8 type diagrams contributing to this decay, as illustrated in Fig.\ref{fig1}.
For the factorizable emission diagrams (a) and (b), operators $O_{1-4,9,10}$ are
$(V-A)(V-A)$ currents, and the operators $O_{5-8}$ have a
structure of $(V-A)(V+A)$, the sum of the their amplitudes are
written as $F_{e\pi}$ and $F_{e\pi}^{P1}$.
In some other cases, we need to do Fierz transformation
for the $(V-A)(V+A)$ operators and get $(S-P)(S+P)$ ones which hold right flavor
and color structure for factorization to work.
The contribution from the operator $(S-P)(S+P)$ type is written as $F_{e\pi}^{P2}$.
Similarly, for the factorizable annihilation diagrams (g) and (h),
the contributions from $(V-A)(V-A), (V-A)(V+A), (S-P)(S+P)$ currents
are
$F_{a\pi}, F_{a\pi}^{P1}$ and $F_{a\pi}^{P2}$.
For the nonfactorizable spectator diagrams (c, d)
and the nonfactorizable annihilation diagrams (e, f), these three kinds of
contributions can be written as $M_{e\pi},
M_{e\pi}^{P1}, M_{e\pi}^{P2}$ and $M_{a\pi}, M_{a\pi}^{P1}, M_{a\pi}^{P2}$, respectively.
Since these amplitudes are similar
to those $B \to f_0(980)K(\pi,\etapp)$ \cite{wwang,zhangzq1} or $B \to a_0(980)K$ \cite{ylshen},
we just need to replace
some corresponding wave functions and parameters.

Combining the contributions from different diagrams, the total decay
amplitudes for these decays can be written as:
\be 2{\cal M}
(\bar{B}^0\to a_0^0\pi^0) &=&
\xi_u\left[(-M_{e\pi}+M_{a\pi}+M_{ea_0}+M_{aa_0})C_2+(F_{a\pi}+F_{ea_0}+F_{aa_0})a_2\right]\nonumber\\
&& -\xi_t \left\{
(M_{e\pi}^{P1}+M_{a\pi}^{P1}+M_{ea_0}^{P1}+M_{aa_0}^{P1})(C_5-\frac{1}{2}C_7)\right.\nonumber\\
&&\left.\left.+
\left(M_{e\pi}+M_{a\pi}+M_{ea_0}+M_{aa_0}\right)(C_3+2C_4-\frac{1}{2}C_9+\frac{1}{2}C_{10})\right.\right.\nonumber\\&&+
\left(M_{e\pi}^{P2}+M_{a\pi}^{P2}+M_{ea_0}^{P2}
+M_{aa_0}^{P2}\right.)
(2C_6+\frac{1}{2}C_8)+\left(F_{a\pi}+F_{ea_0}\right.\nonumber\\&& \left.+F_{aa_0}\right)(2a_3+a_4-2a_5-\frac{1}{2}a_7+\frac{1}{2}a_9-\frac{1}{2}a_{10})
\nonumber\\
&&\left.+\left(F_{e\pi}^{P2}+F_{a\pi}^{P2}+F_{ea_0}^{P2}+F_{aa_0}^{P2}\right)(a_6-\frac{1}{2}a_8)\right\}
, \label{eq:m1}\en
\be {\cal M}
(\bar{B}^0\to a_0^-\pi^+) &=&
\xi_u\left[F_{aa_0}a_2+M_{e\pi}C_1+M_{aa_0}C_2\right]
 -\xi_t \left\{
M_{a\pi}^{P1}\left(C_5-\frac{1}{2}C_7\right)\right.\nonumber\\ &&\left.+M_{e\pi}^{P1}(C_5+C_7)
+M_{a\pi}(C_3+C_4-\frac{1}{2}C_9-\frac{1}{2}C_{10})+M_{aa_0}(C_4+C_{10})\right.\nonumber\\ &&\left.\left.
M_{e\pi}\left(C_3+C_9\right)+
M_{a\pi}^{P2}(C_6-\frac{1}{2}C_8)+M_{aa_0}^{P2}(C_6+C_8)\right.\right.\nonumber\\ &&+
F_{a\pi}(a_3+a_4-a_5+\frac{1}{2}a_7-\frac{1}{2}a_9-\frac{1}{2}a_{10})\nonumber\\&&\left.+F_{aa_0}
(a_3+a_9-a_5-a_7)
+F_{e\pi}^{P2}(a_6+a_8)+F_{a\pi}^{P2}\left(a_6-\frac{1}{2}a_8\right)\right\}
, \label{eq:m2}\en
\be {\cal M}
(\bar{B}^0\to a_0^+\pi^-) &=&
\xi_u\left[F_{ea_0}a_1+F_{a\pi}a_2+M_{ea_0}C_1+M_{a\pi}C_2\right]
 -\xi_t \left\{
M_{aa_0}^{P1}\left(C_5-\frac{1}{2}C_7\right)\right.\nonumber\\ &&\left.+M_{ea_0}^{P1}(C_5+C_7)
+M_{aa_0}(C_3+C_4-\frac{1}{2}C_9-\frac{1}{2}C_{10})+M_{a\pi}(C_4+C_{10})\right.\nonumber\\ &&\left.\left.
M_{ea_0}\left(C_3+C_9\right)+
M_{aa_0}^{P2}(C_6-\frac{1}{2}C_8)+M_{a\pi}^{P2}(C_6+C_8)\right.\right.\nonumber\\ &&+
F_{aa_0}(a_3+a_4-a_5+\frac{1}{2}a_7-\frac{1}{2}a_9-\frac{1}{2}a_{10})+F_{ea_0}(a_4+a_{10})\nonumber\\&&\left.+F_{a\pi}
(a_3+a_9-a_5-a_7)
+F_{ea_0}^{P2}(a_6+a_8)+F_{aa_0}^{P2}\left(a_6-\frac{1}{2}a_8\right)\right\}
, \label{eq:m3}\en
\be \sqrt{2}{\cal M}
(B^-\to a_0^0\pi^-) &=&
\xi_u\left[M_{e\pi}C_2+(-M_{a\pi}+M_{ea_0}+M_{aa_0})C_1+(-F_{a\pi}+F_{ea_0}+F_{aa_0})a_1\right]\nonumber\\
&& -\xi_t \left\{
-M_{e\pi}^{P1}(C_5-\frac{1}{2}C_7)+(-M_{a\pi}^{P1}+M_{ea_0}^{P1}+M_{aa_0}^{P1})(C_5+C_7)\right.\nonumber\\
&&\left.\left.+
M_{e\pi}(-C_3+\frac{1}{2}C_9+\frac{3}{2}C_{10})+\left(-M_{a\pi}+M_{ea_0}+M_{aa_0}\right)(C_3+C_9)\right.\right.\nonumber\\&&
\left.+
\frac{3}{2}C_8M_{e\pi}^{P2}
+\left(-F_{a\pi}+F_{ea_0}+F_{aa_0}\right)(a_4+a_{10})
-F_{e\pi}^{P2}(a_6-\frac{1}{2}a_8)\right.\nonumber\\&& \left.+(-F_{a\pi}^{P2}+F_{ea_0}^{P2}+F_{aa_0}^{P2})(a_6+a_8)\right\}
, \label{eq:m4}\en
\be \sqrt{2}{\cal M}
(B^-\to a_0^-\pi^0) &=&
\xi_u\left[M_{ea_0}C_2+(-M_{aa_0}+M_{e\pi}+M_{a\pi})C_1+F_{ea_0}a_2+(-F_{aa_0}+F_{e\pi}+F_{a\pi})a_1\right]\nonumber\\
&& -\xi_t \left\{
-M_{ea_0}^{P1}(C_5-\frac{1}{2}C_7)+(-M_{aa_0}^{P1}+M_{e\pi}^{P1}+M_{a\pi}^{P1})(C_5+C_7)\right.\nonumber\\
&&\left.\left.+
M_{ea_0}(-C_3+\frac{1}{2}C_9+\frac{3}{2}C_{10})+\left(-M_{aa_0}+M_{e\pi}+M_{a\pi}\right)(C_3+C_9)\right.\right.\nonumber\\&&
\left.+
\frac{3}{2}C_8M_{ea_0}^{P2}
+\left(-F_{aa_0}+F_{e\pi}+F_{a\pi}\right)(a_4+a_{10})
-F_{ea_0}^{P2}(a_6-\frac{1}{2}a_8)\right.\nonumber\\&& \left.+
F_{ea_0}(-a_4-\frac{3}{2}a_7+\frac{3}{2}a_9+\frac{1}{2}a_{10})+(-F_{aa_0}^{P2}+F_{ea_0}^{P2}+F_{a\pi}^{P2})(a_6+a_8)\right\}
, \label{eq:m5}\en
where $\xi_u = V_{ub}V_{ud}^*$, $\xi_t = V_{tb}V_{td}^*$. The
combinations of the Wilson coefficients are defined as usual
\cite{AKL}:
 \be
a_{1}(\mu)&=&C_2(\mu)+\frac{C_1(\mu)}{3}, \quad
a_2(\mu)=C_1(\mu)+\frac{C_2(\mu)}{3},\non
a_i(\mu)&=&C_i(\mu)+\frac{C_{i+1}(\mu)}{3},\quad
i=3,5,7,9,\non
a_i(\mu)&=&C_i(\mu)+\frac{C_{i-1}(\mu)}{3},\quad
i=4, 6, 8, 10.\label{eq:aai} \en

\section{Numerical results and discussions} \label{numer}
\begin{table}
\caption{Input parameters used in the numerical calculation\cite{pdg08}.}\label{para}
\begin{center}
\begin{tabular}{c |cc}
\hline \hline
 Masses &$m_{a_0}=0.9847 \mbox{ GeV}$,   &$ m_0^{\pi}=1.3 \mbox{ GeV}$, \\
  & $ M_B = 5.28 \mbox{ GeV}$,&$ m_{\pi}=0.14 \mbox{ GeV}$,\\
 \hline
  Decay constants &$f_B = 0.19 \mbox{ GeV}$,  & $ f_{\pi} = 0.13
 \mbox{ GeV}$,\\
 \hline
Lifetimes &$\tau_{B^\pm}=1.671\times 10^{-12}\mbox{ s}$, &
$\tau_{B^0}=1.530\times 10^{-12}\mbox{ s}$,\\
 \hline
$CKM$ &$V_{tb}=0.9997$, & $V_{td}=0.0081e^{-i21.6^{\circ}}$,\\
 &$V_{ud}=0.974$, & $V_{ub}=0.00393e^{-i60^{\circ}}$.\\
\hline \hline
\end{tabular}
\end{center}
\end{table}

In the numerical calculation, we will use the input parameters as listed
in Table~\ref{para}.

In the B-rest frame, the decay rates of $B\to a_0(980)\pi$ can be written as:
\be
\Gamma=\frac{G_F^2}{32\pi m_B}|{\cal M}|^2(1-r^2_{a_0})
\en
where $r_{a_0}=m_{a_0}/m_B$ and ${\cal M}$ is the total decay amplitude of
$B\to a_0(980)\pi$, which has been given  in  section \ref{results}.

\begin{table}
\caption{Branching ratios ($\times 10^{-6}$) for the decays $\bar B^0\to a_0^0\pi^0, a_0^\pm\pi^\mp$
 and $B^-\to a_0^-\pi^0, a_0^0\pi^-$. The first theoretical error is from
the the scalar meson decay constant, the second and the third one are Gengebauer moments $B_1$ and $B_3$
for twist-2 LCDAs of $a(980)$. }\label{brch}
\begin{center}
\begin{tabular}{c|c|c|c}
   \hline \hline
   Channel & This work  &Data &QCDF \cite{CCYscalar}  \\
   \hline
    $\bar{B}^0\to a_0^0\pi^0 $ &$0.51^{+0.08+0.09+0.00}_{-0.07-0.09-0.00}$&--&0.2\\
\hline
    $\bar{B}^0\to a_0^+\pi^- $ &$0.86^{+0.10+0.14+0.01}_{-0.09-0.14-0.00}$&--&7.6\\
    $\bar{B}^0\to a_0^-\pi^+ $ &$0.51^{+0.05+0.09+0.07}_{-0.06-0.09-0.06}$&--&0.6\\
    $\bar{B}^0\to a_0^+\pi^-+a_0^-\pi^+ $ &$0.93^{+0.10+0.15+0.02}_{-0.10-0.14-0.00}$&$<3.1$&--\\
\hline
    $B^-\to a_0^-\pi^0$ &$0.41^{+0.00+0.00+0.00}_{-0.13-0.14-0.12}$&$<1.4$&0.2\\
    $B^-\to a_0^0\pi^-$ &$2.8^{+0.00+0.00+0.00}_{-0.79-0.85-0.58}$&$<5.8$&3.4\\
   \hline\hline
\end{tabular}
   \end{center}
\end{table}

Using the wave functions as specified in previous section and the input parameters
listed in Table~\ref{para}, it is straightforward to calculate the
CP-averaged branching ratios for the considered decays, which are listed in Table~\ref{brch}.
In this table, we have included theoretical errors arising from the uncertainties
in the scalar meson
decay constant $\bar f_{a_0}$ and the Gengebauaer moments $B_1$ and $B_3$ for twist-2 LCDAs
of $a_0(980)$.

From the numerical results, one can find that: Firstly, the branching ratio of
$\bar{B}^0\to a_0^0\pi^0$ is larger than that of $\bar{B}^0\to f_0(980)\pi^0$ \cite{zhangzq1},
for the small $u\bar{u}$ and $d\bar{d}$ component in the $f_0(980)$, but much
smaller than the branching ratio $Br(\bar{B}^0\to \pi^0\pi^0)=(1.62\pm0.31)\times 10^{-6}$.
Since the scalar meson $a_0(980)$ has vanishing
decay constants in the isospin limit, the $a_0^-\pi^+$ rate is smaller than that of
its conjugated channel. Secondly, the ratio of these two
rates is about $1/2$. If one includes the interference between these two decay
modes, the branching ratio of $\bar{B}^0\to a_0^-\pi^+ + a_0^+\pi^-$ is close to
$1\times10^{-6}$.
Thirdly, for the two charged decays, it is the same reason with the previous two
neutral ones, and the $a_0^-\pi^0$ rate is much smaller than that of $a_0^0 \pi^-$.
Lastly, the 2-quark model  supposition of $a_0(980)$ can not be ruled out by the
current experimental data, and this point is
different from the QCDF prediction \cite{CCYscalar}.
Since only the upper limits for these channels are available now
and the daughter branching fraction has been taken to be $100\%$,
so the further accurate data are needed to clarify this discrepancy.

\begin{figure}[t,b]
\begin{center}
\includegraphics[scale=0.7]{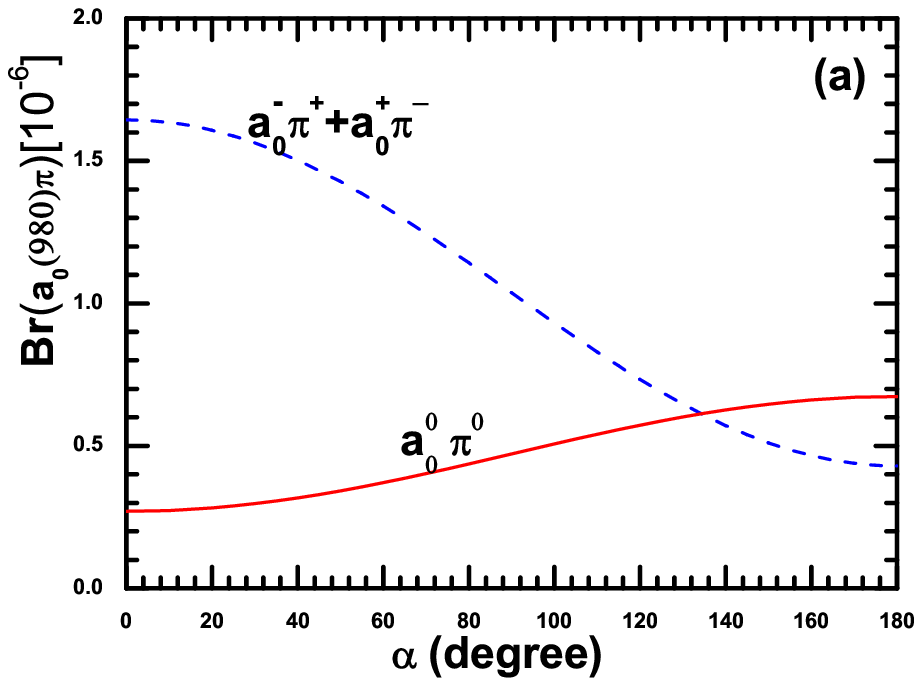}
\includegraphics[scale=0.7]{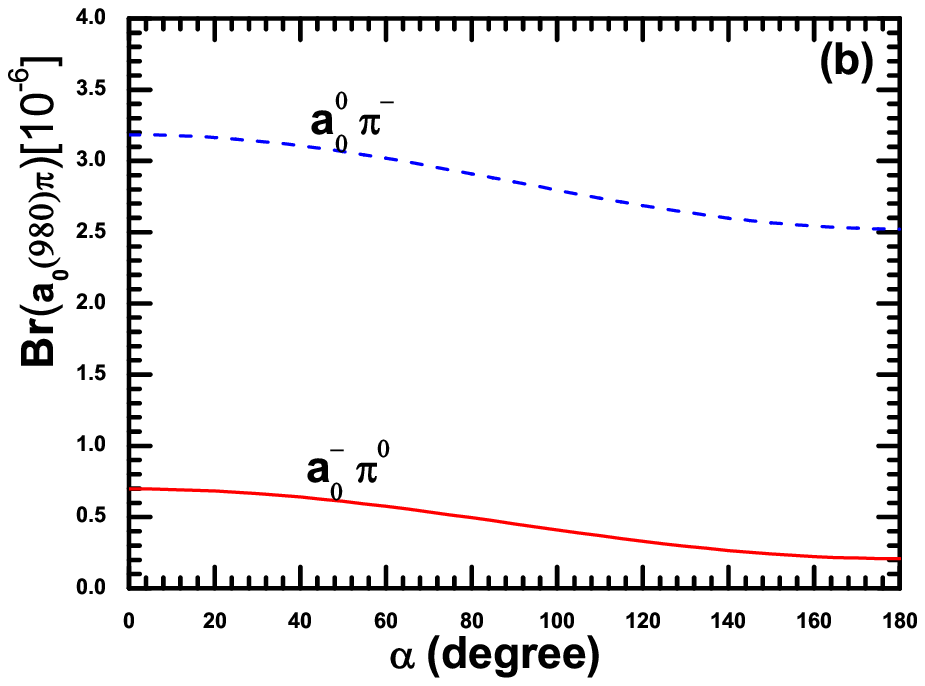}
\vspace{0.3cm} \caption{Branching
ratios (in units of $10^{-6}$) of (a) $\bar{B}^0\to a_0^0\pi^0$ (solid
curve), $\bar{B}^0\to a_0^-\pi^++a_0^+\pi^-$ (dashed curve) and (b) $B^-\to a_0^-\pi^0$
(solid curve), $B^-\to a_0^0\pi^-$ (dashed curve) decays as a function of the CKM angle $\alpha$.}\label{f0pi}
\end{center}
\end{figure}

In Fig.~\ref{f0pi}, we plot the branching ratios of $\bar{B}^0\to a_0^0\pi^0, a_0^-\pi^++a_0^+\pi^-$ and
$B^-\to a_0^0\pi^-, a_0^-\pi^0$
as functions of the CKM angle $\alpha=arg[-\frac{V^*_{td}V_{tb}}{V^*_{ud}V_{ub}}]$. From these figures, it is found that the branching ratio of
$\bar{B}^0\to a_0^-\pi^++a_0^+\pi^-$ is more
sensitive to the angle $\alpha$ than those of other three decay channels. It has been claimed in Ref.\cite{CCYscalar} that the
$B^0-\bar B^0$ interference plays no role in the $a_0^\pm\pi^\mp$ channels. We disagree and we argue that though the $B^0-\bar B^0$
interference in the $a_0^\pm\pi^\mp$ channels may be not as strong as that in some other channels,
such as $a_1(1260)^
\mp\pi^\pm$
modes. There must exist certain interference among them, as can be seen easily in
the numerical values in Table II,
so it may be unreasonable to regard the decays $\bar B^0\to a_0^\mp\pi^\pm$ as
the self-tagging decays in the experiment.


Now we turn to the evaluations of the CP-violating asymmetries of $B^-\to a_0^-\pi^0, a_0^0\pi^-$ and
$\bar B\to a_0^0\pi^0, a_0^\pm\pi^\mp$ decays in pQCD approach. For
the charged decay channels, the direct CP-violating asymmetry can
be defined as:
\be
\acp^{dir}=\frac{ |\overline{\cal A}|^2-|{\cal A}|^2 }{
 |{\cal A}|^2+|\overline{\cal A}|^2}\;.
 \en

For the neutral decays $\bar B^0\to a_0^0\pi^0$,
there are both direct $CP$ asymmetry $A^{dir}_{CP}$ and
mixing-induced $CP$ asymmetry $A^{mix}_{CP}$. The time dependent
$CP$ asymmetry of $B$ decay into a $CP$ eigenstate $f$ $(a_0^0\pi^0)$ is defined
as: \be {\cal A}_{CP}(t)={\cal A}^{dir}_{CP}(B_d\to f)\cos(\Delta M
t)+{\cal A}^{mix}_{CP}(B_d\to f)\sin(\Delta M t),\en with \be {\cal
A} ^{dir}_{CP}(B_d\to f)&=&\frac{|\lambda|^2-1}{1+|\lambda|^2}\;,
\;\;\;
{\cal A}^{mix}_{CP}(B_d\to f)=\frac{2 Im \lambda}{1+|\lambda|^2}\;,\\
\lambda&=&\eta e^{-2i\beta}\frac{{\cal A}(\bar B_d\to f)}{{\cal
A}(B_d\to f)}\;,\en where  $\eta=\pm1$ depends on the $CP$ eigenvalue
of $f$, $\Delta M$ is the mass difference of the two neutral $B$
meson eigenstates.

\begin{figure}
\begin{center}
\includegraphics[scale=0.65]{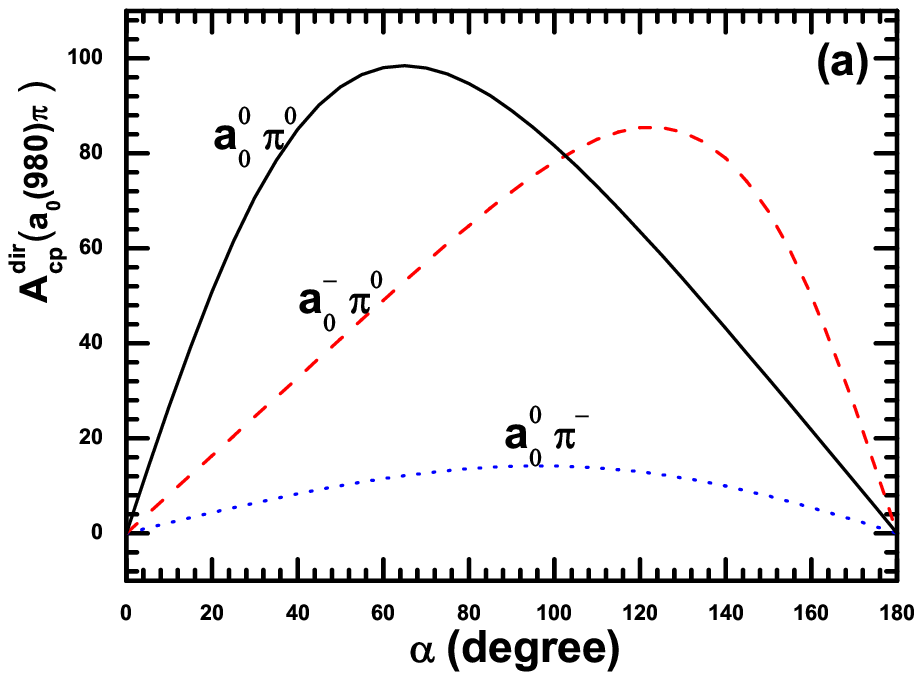}
\includegraphics[scale=0.65]{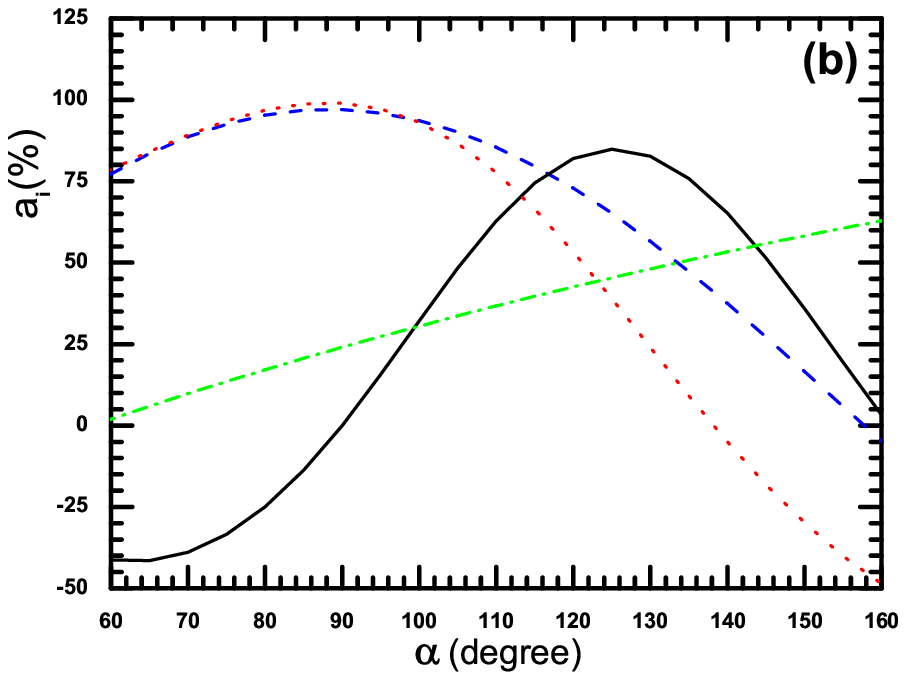}
\vspace{0.3cm} \caption{The direct CP asymmetries (a) of the decays
$\bar{B}^0\to a_0^0\pi^0$ (solid curve), $B^-\to a_0^-\pi^0$ (dashed curve), $B^-\to a_0^+\pi^-$ (dotted cureve)
and the CP asymmetry parameters (b) of the decay $\bar{B}^0\to a_0^+\pi^-+a_0^-\pi^+$: $a_{\epsilon'}$ (dash-dotted curve)
, $a_{\overline \epsilon'}$ (dotted cureve),
$a_{\epsilon+\epsilon'}$ (dashed cureve) and $a_{\epsilon+\overline
\epsilon'}$ (solid curve) as functions of the CKM angle
$\alpha$. }\label{pidir}
\end{center}
\end{figure}

As to the decays $\bar{B}\to a_0^\pm\pi^\mp$, since both $B^0$ and $\bar B^0$ can decay into the final state $a_0^+\pi^-$
and $a_0^-\pi^+$, the four time-dependent decay widths for $B^0(t)\to a_0^+\pi^-, \bar{B}^0(t)\to a_0^-\pi^+,
B^0(t)\to a_0^-\pi^+$ and $\bar{B}^0(t)\to a_0^+\pi^-$ can be expressed by four basic matrix elements:
\be
 g= \langle a_0^+\pi^- |H_{eff}| B^0\rangle ,\; \;\; \;h=\langle a_0^+\pi^- |H_{eff}|
 \bar{B}^0 \rangle,
\non \bar g= \langle a_0^-\pi^+ |H_{eff}| \bar B^0\rangle
,\;\;\; \; \bar h=\langle a_0^-\pi^+ |H_{eff}|
  B^0\rangle.
 \en

Following the notation of Ref.\cite{lvcd,zhangzq2}, the four CP violating parameters are given by the following formulae:
\be
 a_{\epsilon'}&=&\frac{|g|^2-|h|^2}{|g|^2+|h|^2},\quad
 a_{\epsilon+\epsilon'}=\frac{-2Im(\frac{q}{p}\frac{h}{g})}{1+|h/g|^2} \non
 a_{\bar \epsilon'}&=&\frac{|\bar h|^2-|\bar g|^2}{|\bar h|^2+|\bar g|^2},\quad
 a_{\epsilon+\bar \epsilon'}=\frac{-2Im(\frac{q}{p}\frac{\bar g}{\bar h})}{1+|\bar g/\bar
 h|^2}, \label{eq:a41}
 \en
where $q/p=e^{-2i\beta}$ is defined via
\be
 B_1=p|B^0\rangle+q| \overline B^0\rangle,\;\; \; \;B_2=p|B^0\rangle-q| \overline
 B^0\rangle,
\en
with $|p|^2 + |q|^2 = 1$ and $\beta$ being a CKM angle.

From the Fig.\ref{pidir}(b), one can find the central values of the CP-violation parameters:
\be
a_{\epsilon'}=0.31, \quad a_{\epsilon+\epsilon'}=0.94,\quad
a_{\bar \epsilon'}=0.93, \quad a_{\epsilon+\bar \epsilon'}=0.32;
\en
for $\alpha=100^\circ$.

The direct CP asymmetries are shown in Fig.\ref{pidir}(a).
The branching ratio of decay $B^-\to a_0^0\pi^-$ is the
largest one among the considered channels,
while its direct CP asymmetry is the smallest, $14\%$.
Although the CP asymmetries for the decays $\bar{B}^0\to a^0_0(980)\pi^0
, B^-\to a^-_0(980)\pi^0$
are large, about $(70\thicksim80)\%$, it is still difficult to measure them,
since their branching ratios are small, around $(4\thicksim5)\times 10^{-7}$.


\section{Conclusion}\label{summary}

In this paper, we calculate the branching ratios and CP-violating
asymmetries of $ \bar{B}^0\to a_0^0\pi^0$, $\bar{B}^0\to a_0^+\pi^-$,
$\bar{B}^0\to a_0^-\pi^+$, $B^-\to a_0^0\pi^-$ and $B^-\to a_0^-\pi^0$ decays
in the pQCD factorization approach by identifying $a_0(980)$ as the
2-quark content.
Using the decay constants and light-cone distribution amplitudes
derived from QCD sum-rule method, we find that:
\begin{itemize}
\item
Since the scalar meson $a_0(980)$ has vanishing decay constant in the isospin limit,
one can find $Br(\bar{B}^0\to a_0^+\pi^-)>Br(\bar{B}^0\to a_0^-\pi^+)$ and
$Br(B^-\to a_0^0\pi^-)>Br(B^-\to a_0^-\pi^0)$.

\item
If one includes the $B^0-\bar B^0$ interference in the $a_0^\pm\pi^\mp$ channels,
$Br(\bar{B}^0\to a_0^+\pi^-+a_0^-\pi^+)=0.93
\times 10^{-6}$, which is close to $Br(\bar{B}^0\to a_0^+\pi^-)=0.86\times 10^{-6}$.
There must exist certain
interference in these two neutral channels.

\item
Although the CP asymmetries of
$\bar{B}^0\to a^0_0(980)\pi^0, B^-\to a^-_0(980)\pi^0$
are large, about $(70\thicksim80)\%$, it is still difficult to measure them, since
their branching ratios are small.

\item
For our predictions agree well with the currently available experimental upper limits,
the 2-quark model supposition for $a_0(980)$ can not be ruled out. This point of
view is different from the conclusion obtained by using the QCDF approach.

\end{itemize}

\section*{Acknowledgment}
Z.Q.~Zhang would like to thank H.Y.~Cheng and W.~Wang for helpful  discussions.
This work is partly supported  by the National Natural Science
Foundation of China under Grant No.10575052 and 10735080.


\end{document}